\documentclass[12pt]{iopart}
\usepackage{graphicx}
\usepackage{siunitx}
\DeclareRobustCommand{\vett}[1]{
  \ifcat#1\relax
    \boldsymbol{#1}
  \else
    \mathbf{#1}
  \fi}
\newcommand{\de}{{\rm d}}

\begin{document}

\title[Population Distribution in the Wake of a Sphere]{Population Distribution in the Wake of a Sphere}

\author{Taraprasad Bhowmick$^{1,2,*}$, Yong Wang$^{2,*}$, Michele Iovieno$^3$, Gholamhossein Bagheri$^{2,*}$ and Eberhard Bodenschatz$^{2,*}$}

\address{
$^{1}$ Department of Applied Science and Technology, Politecnico di Torino, Torino, Italy\\
$^{2}$ Laboratory for Fluid Physics, Pattern Formation and Biocomplexity, Max Planck Institute for Dynamics and Self-Organization, G\"ottingen, Germany\\
$^{3}$ Department of Mechanical and Aerospace Engineering, Politecnico di Torino, Torino, Italy
}
\address{$^*$ Am Fa{\ss}berg 17, 37077 G\"ottingen, Germany}
\ead{yong.wang@ds.mpg.de; gholamhossein.bagheri@ds.mpg.de}
\vspace{10pt}
\begin{indented}
\item[]August 2017
\end{indented}

\begin{abstract}
The fluid physics of the heat and mass transfer from an object in its wake has much importance for natural phenomena as well as for many engineering applications. Here, we report numerical results on the population density of the spatial distribution of fluid velocity, pressure, scalar concentration and scalar fluxes of a wake flow past a sphere in the steady wake regime (Reynolds number 25 to 285). We find the population density to be well described by a Lorentzian distribution.
We observe this apparently universal form both in the symmetric wake regime and in the more complex three dimensional wake structure of the steady oblique regime with Reynolds number larger than 225.
The population density distribution identifies the increase in dimensionless kinetic energy and scalar fluxes with the increase in Reynolds number, whereas the dimensionless scalar population density shows negligible variation with the Reynolds number.
\end{abstract}

%
\noindent{\it Keywords}: Sphere wake; steady flow; axisymmetric wake; oblique wake; lattice Boltzmann method; direct numerical simulation; population density; Lorentzian distribution.
%
%
%
%

\section{Introduction}

The interactions between spherical bodies, such as, particles, bubbles and drops, and the ambient through which they move is a vast area of research, which has attracted attention over centuries in various scientific disciplines \cite{Clift1978,Michaelides2006}.
The flow past a sphere presents different regimes at different Reynolds number $Re$.
The steady axisymmetric structure of a wake at low Reynolds number, up to $Re\sim210$ \cite{Taneda1956,Natarajan1993,Tomboulides2000}, followed by a steady oblique wake structure up to $Re\sim280$ \cite{Johnson1999,Magarvey1961}, and an unsteady structure of the wake at higher $Re$ \cite[etc.]{Fornberg1988,Ormieres1999} had been studied both experimentally and numerically.
The drag coefficient $C_D$ of a sphere, which varies with the roughness of the sphere surface and $Re$, was studied in detail both in experiments \cite{Bodenschatz2011,Roos1971,Eichhorn1964,Unnikrishnan1991} and numerical simulations \cite[etc.]{Tabata1998,Birouk2007}.
At present, how the drag, lift and pressure coefficients vary both locally as well as globally with respect to the sphere is well known \cite{Tomboulides2000,Birouk2007,Wu2012}.
The two dimensional structures of the streamlines, vorticity and pressure contours along the orthogonal central planes through the sphere are also well known over various studies \cite{Tomboulides2000,Johnson1999,Bagchi2000}.

For many engineering applications and natural processes, the interaction between a sphere and the ambient also involves transport of various scalar species, either passively advected by the ambient flow or interacting actively with the flow through various physical processes, for example, through evaporation, buoyancy.
The rate of scalar transport, in particular the convective heat transfer from spherical objects at various $Re$, has been investigated both numerically \cite{Bagchi2000,Richter2012} and experimentally \cite{Kramers1946,Gibson1968,Yuge1960,Will2017} to determine the heat transfer coefficient.
Similar to the drag coefficient, attention was given to the dependence of the local Nusselt number (a ratio of the convective and the diffusive (conductive) heat transfer) on the sphere surface and its global average for different $Re$ \cite{Bagchi2000,Will2017}. 
The profiles of the dimensionless temperature contours along the central orthogonal plane for various $Re$ have also been described in the literature \cite{Bagchi2000,Chouippe2019}.
A coupled system involving an interplay between different scalars can also be present, for example, in case of the phase change during droplets evaporation or freezing resulting in heat and mass exchange with the ambient air.
Such interaction has also been studied both experimentally \cite{Ranz1952,Friedlander1957} and numerically \cite{Dennis1973,Chouippe2019}.
All these studies are mainly concerned with the average scalar flux at the surface of the sphere, which determines the mass and temperature change rate of the sphere.
An overall description on the spatial structure of the wake, including the scalar concentrations and the convective fluxes for various $Re$ was not fully explored.

Descriptive statistics on the spatial structure of the wake is of primary importance if the extent of the wake with certain properties needs to be quantified.
Supersaturation in the wake of a precipitating cloud water droplets \cite{Bhowmick2020,Chouippe2020} for example requires a detailed analysis of the wake population.
In this paper, we present a comprehensive numerical study on the details of the momentum and scalar transport in the wake of a sphere using a population density distribution for the steady axisymmetric and oblique wake regimes.
A brief introduction of the numerical methods and computational details are described in Section \ref{Numerical Method and Simulation Setup}. Results are presented and discussed in Section \ref{Results}. Finally conclusions are given in Section \ref{Conclusion}. 

\section{Physical Model, Numerical Method and Boundary Conditions} \label{Numerical Method and Simulation Setup}

We consider the flow which develops past a sphere, placed in incompressible viscous fluid with velocity $\vett{u}_\infty=(u_\infty,0,0)$, pressure $p_\infty$, and a constant density $\rho$. Together with the balances of mass and momentum, we consider also the transport of a passive scalar, that is any contaminant present in low concentration so that it does not influence the flow. Such dynamics is described in an Eulerian framework by an advection–diffusion (AD) equation.
If $d_s$ is the diameter of the sphere, $\theta$ the passive scalar concentration, $\theta_s$ and $\theta_\infty$ are the scalar concentration on the surface of the sphere and in the external flow respectively, the problem can be suitably made dimensionless by using $d_p$, $u_\infty$ and $\theta_s-\theta_\infty$ as scales, and therefore by defining the dimensionless position, time, velocity, pressure, and scalar concentration as,

\begin{equation*}
 \vett{x}^* = \frac{\vett{x}}{d_s},
 \;\;
 t^* = \frac{t u_\infty}{d_s},
 \;\;
 \vett{u}^* = \frac{\vett{u}}{u_\infty},
 \;\;
 p^* = \frac{p-p_\infty}{\rho u_\infty^2},
 \;\;
 \theta^* = \frac{\theta-\theta_\infty}{\theta_s-\theta_\infty}.
\end{equation*}

Therefore, the dimensionless incompressible Navier-Stokes (NS) equations and the one-way coupled AD equation for the scalar are,
\begin{eqnarray}
\centering
\nabla^*\cdot\vett{u}^*  &=& 0 , \label{eq.cont_ND} \\
\frac{\partial\vett{u}^*}{\partial t^*}+\vett{u}^*\cdot\nabla^*\vett{u}^* &=& -\nabla^* p^* +\frac{1}{Re} \nabla^{*2} \vett{u}^* , \label{eq.qdm_ND}\\
\frac{\partial \theta^*}{\partial t^*}+\vett{u}^*\cdot\nabla^* \theta^* &=& \frac{1}{Re\, Sc}\nabla^{*2} \theta^* \label{eq.T_ND},
\end{eqnarray}
where $Re=u_\infty d_s/\nu$ is the Reynolds number ($\nu$ is the kinematic viscosity) and  $Sc=\nu/\kappa_\theta$ is the Schmidt number, ratio between the kinematic viscosity and the scalar diffusivity $\kappa_\theta$. These equations are complemented by uniform flow boundary conditions far from the sphere ($\vett{u}^*\rightarrow (1,0,0)$, $\theta^*\rightarrow 0$) and no slip boundary conditions on the surface of the sphere with a constant scalar concentration ($\vett{u}^*=0$, $\theta^*=1$). For sake of clarity, the $*$ will be omitted in the following.

These governing equations are numerically solved with the lattice Boltzmann method (LBM) \cite{succi2011,Kruger2017}. A code is developed based on the open-source library, Palabos \cite{Latt2020}. In LBM, the particle distribution function $f(\vett{x},t)$ is governed by 
\begin{eqnarray}
\fl f_i(\vett{x}+\vett{c}_i\Delta t,t+\Delta t) = f_i(\vett{x},t) + \Omega_i(\vett{x},t), \; \; \; \; \; \; \Omega_i(\vett{x},t) &= -\frac{\Delta t}{\tau} (f_i(\vett{x},t)-f_i^{eq}(\vett{x},t)). \label{eq.lbm}
\end{eqnarray}
Here $i$ is the index of the discrete velocity $\vett{c}$, which defines the structure of lattice; $\vett{x}$ and $t$ are the location of a lattice node and the time respectively. 
The collision operator $\Omega_i(\vett{x},t)$ models the redistribution of the particle populations at each lattice node.
In this study, we consider the Bhatnagar-Gross-Krook (BGK) collision operator \cite{Qian1992}, with which the population $f_i(\vett{x},t)$ relaxes towards its equilibrium state $f_i^{eq}(\vett{x},t)$ according to the relaxation time scale $\tau$.  $f_i^{eq}(\vett{x},t)$ and $\tau$ are defined as,
\begin{eqnarray}
f_i^{eq}(\vett{x},t) = w_i\rho (1+\frac{\vett{c}_i\cdot\vett{u}}{c_s^2}+\frac{(\vett{c}_i\cdot\vett{u})^2}{2 c_s^4}+\frac{\vett{u}\cdot\vett{u}}{2 c_s^2}) , \; \; \; \; \; \; \nu = c_s^2(\tau-\frac{\Delta t}{2}). \nonumber
\end{eqnarray}
Here $w_i$ is the weight; $c_s$ is the speed of sound. The macroscopic quantities, such as the density $\rho$ and velocity $\vett{u}$ are moments of $f_i(\vett{x},t)$, according to $\rho = \sum_{i}f_i(\vett{x},t) = \sum_{i}f_i^{eq}(\vett{x},t)$ and $\rho\vett{u} = \sum_{i}\vett{c}_i f_i(\vett{x},t) = \sum_{i}\vett{c}_i f_i^{eq}(\vett{x},t)$ respectively. For solving the fluid velocity field, the $D3Q19$ lattice is chosen as the non-linear momentum advection corrections are not very significant in the steady axisymmetric or oblique wake flows \cite{Silva2012}.

The one-way coupling between the fluid momentum $\rho\vett{u}$ and the scalar concentration $\theta$ is solved by another LBM equation similar to equation \ref{eq.lbm}, but with a distribution function $g_i(\vett{x},t)$ for the scalar. To recover the AD equation, the equilibrium distribution function $g_i^{eq}(\vett{x},t)$ \cite{Guo2002a} and the relaxation time scale $\tau_g$ are given as,
\begin{eqnarray}
g_i^{eq}(\vett{x},t) = w_i \theta (1+\frac{\vett{c}_i\cdot\vett{u}}{c_s^2})
, \; \; \; \; \; \; \kappa_\theta = c_s^2(\tau_g-\frac{\Delta t}{2}). \nonumber
\end{eqnarray}
The scalar concentration $\theta$ is calculated according to $\theta = \sum_{i}g_i(\vett{x},t) = \sum_{i}g_i^{eq}(\vett{x},t)$. Since only the zeroth and the first order moments of $g_i(\vett{x},t)$ are used to recover the AD equation from the LBM equation, a $D3Q7$ lattice is used for the scalar field \cite{Kruger2017}.

\begin{figure}
\centering
\includegraphics[width=\textwidth]{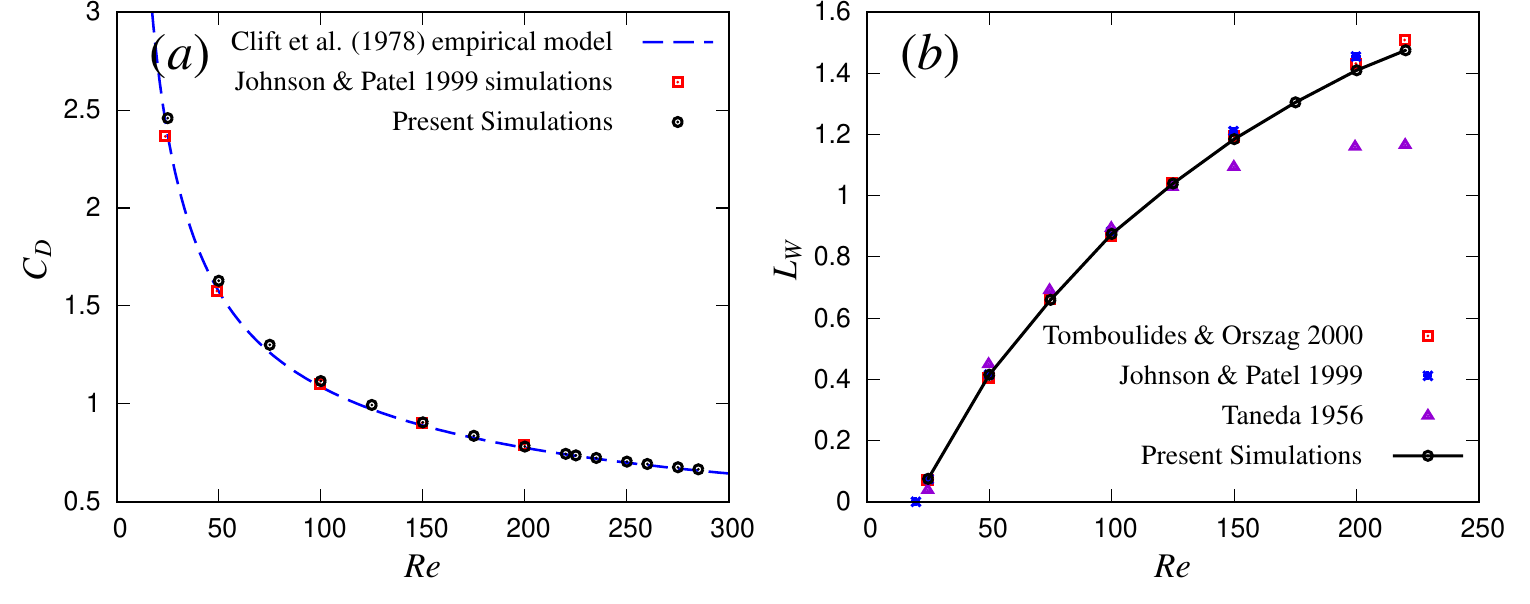}
\caption{(\textit{a}) Drag coefficient $C_D$ and (\textit{b}) wake length $L_W$ normalized with sphere diameter $d_s$ for various steady axisymmetric and oblique $Re$ with existing researches \cite{Clift1978,Johnson1999,Tomboulides2000,Taneda1956}.
}
\label{fig:Fig1}
\end{figure}

The sphere is set in the origin of the reference frame, and the dimensionless domain is $[-5,20]\times[-3.5,3.5]\times[-3.5,3.5]$ in size (5 diameters upstream, 20 diameters downstream and 7 diameters in the transversal directions).
The domain is discretized with a uniform Cartesian mesh with a grid size equal to $1/32$ of the sphere diameter.
Dirichlet and Neumann boundary conditions are considered for the inlet and outlet boundaries, respectively. For the lateral boundaries in transversal directions, periodic boundary conditions are applied. A second order extrapolation scheme, proposed Guo et al. (2002) \cite{Guo2002}, is adopted for the curved boundary of the sphere.

The numerical setup is validated by comparing the drag coefficient, the length of the recirculating zone and the angle of separation with existing researches for the fluid velocity field. Tests have shown the mesh and domain independence for the flow around the sphere in the range of parameters considered.
For example, in Figure \ref{fig:Fig1}(\textit{a}), the drag coefficient $C_D$ obtained from our simulation is compared with empirical equations (equations \ref{eq.CGW78a} and \ref{eq.CGW78b}) of Clift et al. (1978) \cite{Clift1978} and with the numerical results of Johnson and Patel (1999) \cite{Johnson1999}.
The drag coefficient deviates from the empirical equations maximum at $Re=25$, with relative error $3.5\%$, which is further reduced with higher $Re$, e.g. less than $1\%$ at $Re=200$.
Figure \ref{fig:Fig1}(\textit{b}) presents the results of wake length $L_W$ along with numerical results of Johnson and Patel (1999) \cite{Johnson1999}, Tomboulides and Orszag (2000) \cite{Tomboulides2000}, and experimental data of Taneda (1956) \cite{Taneda1956}, which reported transition to unsteady wake for $Re\ge130$.
The scalar field is validated by comparing the normalized scalar profiles with other numerical simulations, which for example shows a maximum of 2 lattice node difference from the temperature profiles of Chouippe et al. (2019) \cite{Chouippe2019} at a similar scalar diffusivity of $Sc=0.7$ (not shown here).

\begin{figure}
\centering
\includegraphics[width=\textwidth]{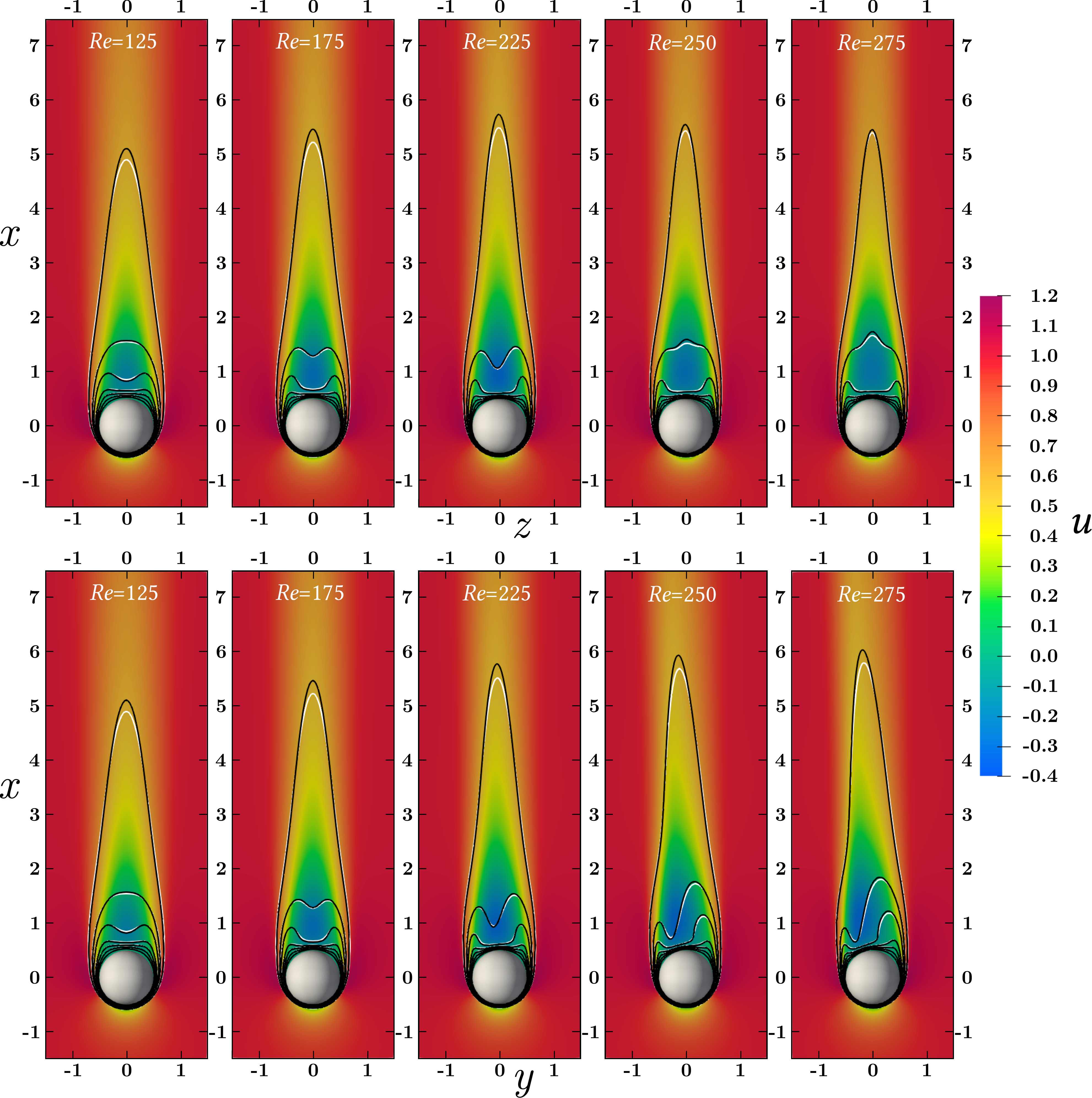}
\caption{Spatial distribution of the dimensionless streamwise component of fluid velocity $u$ in color and the contour lines of a scalar $\theta_1$ in black ($Sc=0.71$) and another scalar $\theta_2$ in white ($Sc=0.61$) for various steady axisymmetric and oblique $Re$.
The visualization is across two central orthogonal planes ($z,x$) and ($y,x$) passing through the center of the sphere with an extent of [-1.5,1.5] along the horizontal axes and [-1.5,7.5] along the vertical $x$ axis.
Contour lines for $\theta_1$ and $\theta_2$ are plotted at magnitudes of $0.2,0.35,0.45,0.6,0.7,0.8$ and $0.9$, ascending from the ambient towards the sphere.
}
\label{fig:Fig2}
\end{figure}

\begin{eqnarray}
\fl C_D &=& \frac{24}{Re}(1 + 0.1935\cdot Re^{0.6305}), \quad
 \rm{if~} 20\le Re\le260. \label{eq.CGW78a} \\
\fl \log_{10}C_D &=& 1.6435 - 1.1242\cdot \log_{10}Re + 0.1558\cdot (\log_{10}Re)^2,
 \quad \rm{if~} 260\le Re\le 1500. \label{eq.CGW78b}
\end{eqnarray}

\section{Results on Spatial Structure of Steady Wake} \label{Results}

Our work focuses on the wake behind a wet sphere in the steady axisymmetric regime ($0\le Re\le 220$) and the steady oblique regime ($225\le Re\le 285$).
The difference in the overall features of these regimes can be appreciated from Figure \ref{fig:Fig2}, which visualizes the streamwise velocity $u$ together with the contours of two advected scalar fields $\theta_1$ and $\theta_2$ of different scalar diffusivities in two perpendicular planes ($z,x$) and ($y,x$) passing through the center of sphere in parallel to the incoming flow. The Schmidt numbers for the scalars are 0.71 and 0.61, respectively, which correspond to the diffusivities of temperature and water vapor in air.
The increase in $Re$ features the thinning of the boundary layer, as well as a shrinking in the lateral extent of the wake and a stretching in the streamwise direction as in Figure \ref{fig:Fig2} up to $Re=220$.
In the oblique regime, a tilt from the centerline ($y=z=0$) along the ($y,x$) plane is observed, which is symmetric along ($z,x$) plane, see also \cite{Johnson1999,Chouippe2019}.
This tilt in the oblique regime increases with $Re$ until the wake becomes unstable and starts shedding vortices at $Re\ge290$.
The apparent decrease in the streamwise length of the wake in the top panel of Figure \ref{fig:Fig2} from $Re=225$ to $275$ is attributed to the tilting of the wake.
The transport of any  scalar $\theta$ is described by the same equation (\ref{eq.T_ND}). The only difference lays in their Schmidt numbers, which governs their relative diffusivity.
The different diffusivities govern the profiles of the scalars at the intermediate values of the dimensionless concentration, which shows differ in the external part away from the sphere boundary and in the far wake (for $\theta\sim 0.2$ to $0.4$), as shown in Figure \ref{fig:Fig2}.
Due to the Schmidt numbers, the gradient of $\theta_1$ is less steep than the gradient of $\theta_2$.
This feature, however, becomes less distinctive at higher concentrations of $\theta_1$ and $\theta_2$ near the sphere surface.

\begin{figure}
\centering
\includegraphics[width=\textwidth]{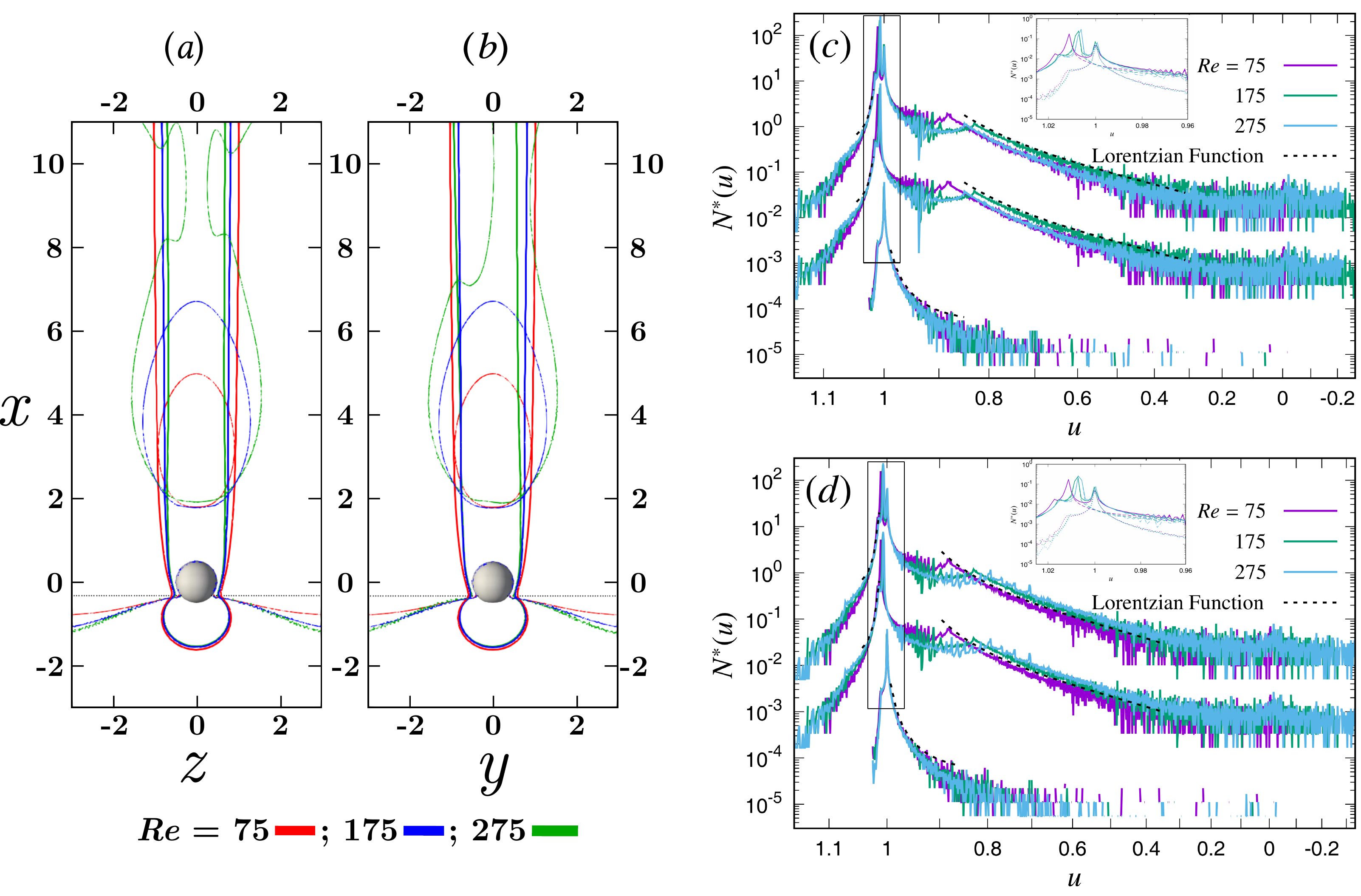}
\caption{Distribution of the dimensionless streamwise velocity component $u$ for various $Re$. $u=0.95$ contours are drawn in solid lines along with $p=0$ pressure contours in dashed thin lines along the orthogonal (\textit{a}) ($z,x$) and (\textit{b}) ($y,x$) planes.
A horizontal dotted line at $x=-0.325$ is drawn to divide the upstream spatial structure of $u$ from the downstream one.
Normalized population density function $N^*(u)=N(u)/A$ ($A$ is the area of the orthogonal plane) for the $u$ sample population across the orthogonal ($z,x$) and ($y,x$) planes are plotted respectively in (\textit{c}) and (\textit{d}).
$N^*(u)$ for the upstream, downstream and the entire planes are respectively plotted as the bottom, middle and the top sets of curves.
A scale difference is created by amplifying the $N^*(u)$ of the downstream and the entire domain 30 and 900 times respectively.
Sample extent in (\textit{a,b}) is $[-3,3]$ along the horizontal and $[-3,11]$ along the vertical axes, whereas, in (\textit{c,d}) it is $[-3.5,3.5]$ along the horizontal and $[-5,20]$ along the vertical axes.
}
\label{fig:Fig3a}
\end{figure}

In order to provide a synthetic description of the flow field, we use a population density approach. For any variable, such as the longitudinal velocity component $u$, its population density distribution $N(u)$ at a $u_0$ magnitude is defined as $N(u_0) = \de V_u(u_0)/\de u$, where $V_u(u_0)$ is the volume of the region in which $u$ is lower that $u_0$.
The distribution of $u$ is shown in Figure \ref{fig:Fig3a} for three different Reynolds numbers ($Re=$75, 175 and 275).
Figure \ref{fig:Fig3a}(\textit{a}) and (\textit{b}) present the contour lines of $u=0.95$ in solid lines and of pressure $p=0$ in dashed thin lines across the ($z,x$) and ($y,x$) orthogonal planes respectively.
The domain can be divided into two main parts: an upstream zone where the flow approaches the sphere 
and a downstream zone dominated by the presence of the wake. 
The dotted horizontal black line in panels (\textit{a}) and (\textit{b}) of Figure \ref{fig:Fig3a}, located at $x=-0.325$, intersects the sphere where the dimensionless pressure $p$ changes sign and distinguishes the two regions.
the velocity component $u$ in the upstream zone ($p\ge0$) does not show significant changes with $Re$, but the above mentioned lateral thinning is visible in the downstream zone, which has mostly negative $p$.
Tilting is also observed in Figure \ref{fig:Fig3a}(\textit{b}) for $Re=275$.

Figure \ref{fig:Fig3a}(\textit{c}) and (\textit{d}) present the population density distribution $N(u)$ of the longitudinal velocity component $u$ in these two zones, computed along the entire orthogonal ($z,x$) and ($y,x$) planes of the computational domain, respectively.
The distribution has been determined by dividing the range of $u$ in 1000 bins, a resolution which allows for a smooth sample distribution while preserving its trend.
In the upstream zone (bottom sets of curves in Figure \ref{fig:Fig3a}(\textit{c,d})), $N(u)$ shows a sharp decrease in population density as $u$ decreases from the external ambient value of $1$  towards the no-slip zero boundary condition at the sphere surface following a Lorentzian function in equation \ref{eq.Lorentz}.
Some sample population with $u\ge1$ is also observed which resembles the region of highest velocity magnitudes near the $p\sim0$ contour line.
In order to create a visible scale separation, the $N(u)$ of the downstream zone is shifted for the middle set of curves in Figure \ref{fig:Fig3a}(\textit{c,d}).
Negative values of velocity identify the recirculation zone behind the sphere.
A large extent of the simulated wake can be well fitted by a Lorentzian distribution.
The crescent like trend right after the ambient $u=1$ is a result of the finite size of the simulation domain.
Similar to the $N(u)$ of the upstream zone, some sample population with $u\ge1$ is also observed in this downstream distribution, which are also coming from the $p\sim0$ region.
$N(u)$ of the entire plane is shifted for the top sets of curves in Figure \ref{fig:Fig3a}(\textit{c,d}) with an amplification of its original magnitudes.
As plotted in the insets, the two highest peaks at $u\sim1$ of the entire plane are the individual contributions from both the upstream and the downstream populations.

The Lorentzian or Lorentz-Cauchy distribution $y(u;A,u_c,b,y_0)$ is a single peak bell-shaped curve, defined as
\begin{eqnarray}
 y(u;A,u_c,b,y_0) &=& y_0 + 2 \; \frac{A}{\pi b} \; \frac{b^2}{4(u-u_c)^2+b^2}, \label{eq.Lorentz}
\end{eqnarray}
where $y(u;A,u_c,b,y_0)$ is the population density of samples of variable $u$, $A$ is its integral over all possible values of $u$, $u_c$ is the position of its maximum where $y$ takes the value $2A/(\pi b)$, with $b$ being the width between its half maximums. Parameter 
$y_0$ is just an offset value.
In the distribution of  $u$, Figure \ref{fig:Fig3a}(\textit{c,d}), a Lorentzian trend is observed in the intermediate range, which corresponds to the boundary layer and to the region external to the wake. An increase in $N(u)$ is observed with increasing Reynolds numbers, indicating an increase in the dimensionless kinetic energy in this region. 
The out of plane tilting induced by the oblique wake at $Re=275$ produces small spikes on top of an overall Lorentzian trend of the sample population along the ($y,x$) plane, as seen in Figure \ref{fig:Fig3a}(\textit{d}).
However, the oblique wake regime retains a symmetric structure along the ($z,x$) plane in our simulations but the out of plane tilting impacts the sample population. Therefore, $N(u)$ in Figure \ref{fig:Fig3a}(\textit{c}) for $Re=275$ only indicates a lower yet a smooth Lorentzian trend.

The existence of such a trend in the distribution of a variable indicates the existence of a matching region where the variable shows an algebraic variation from the values in the wake to the values in the external ambient. If the flow is axisymmetric and the flow structures are elongated in the streamwise direction, this variation is in the radial direction proportional to $(y^2+z^2)^{-1}$ (inverse of the square of the lateral distance from the axis).
This algebraic matching region is not only present in the velocity field, but also in the associated pressure field and in the passively transported scalars.

\begin{figure}
\centering
\includegraphics[width=\textwidth]{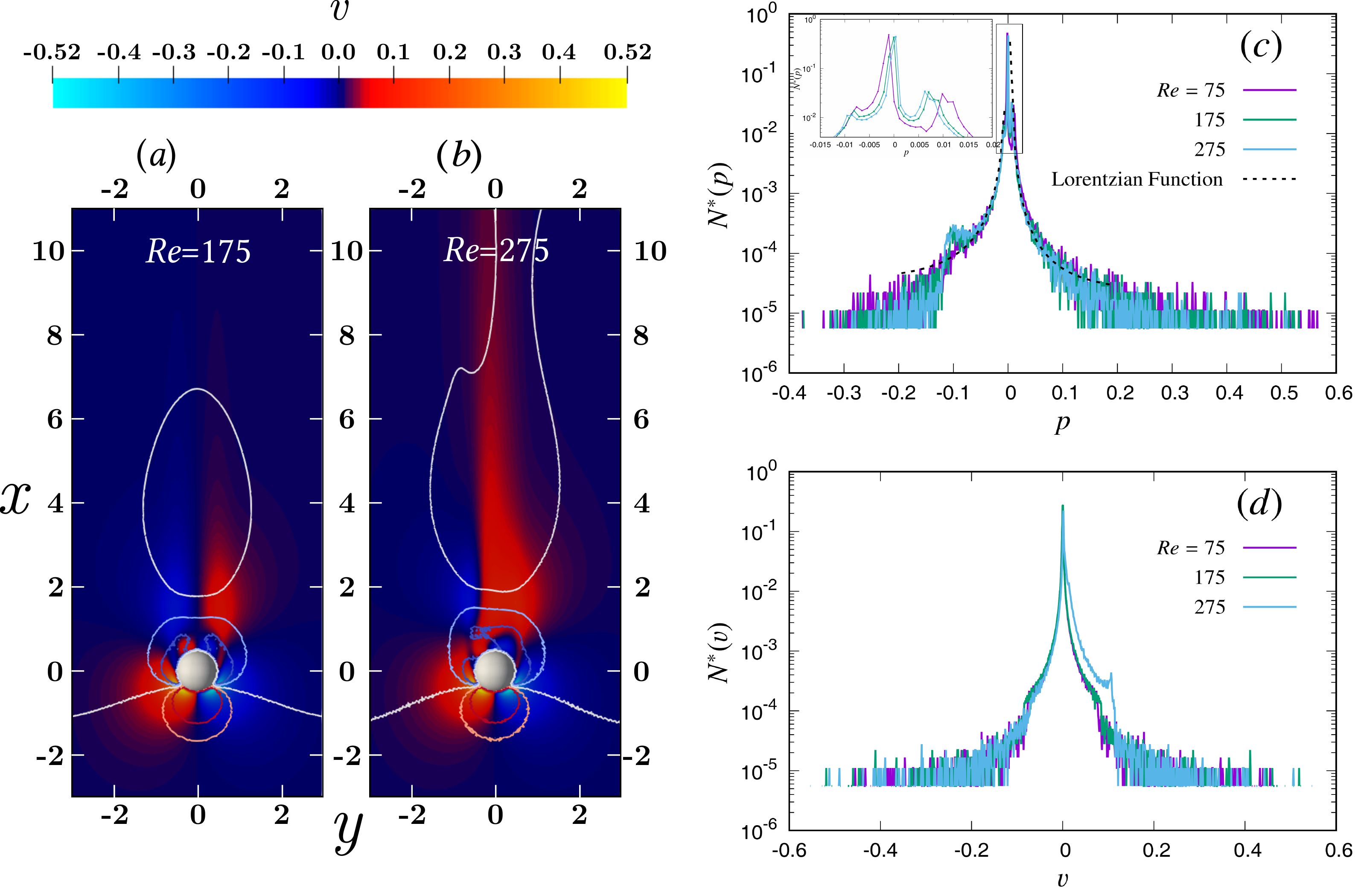}
\caption{Distribution of the pressure $p$ and velocity component $v$ for various $Re$. Spatial distribution of $v$ in color along with the contour lines of $p$ at $0.1,0.05,0.0,-0.05,-0.1$ magnitudes respectively in red, orange, white, cyan, and blue solid lines along the orthogonal ($y,x$) plane for the axisymmetric $Re=175$ in (\textit{a}) and for the oblique $Re=275$ in (\textit{b}).
Normalized population density of pressure as $N^*(p)$ across the entire orthogonal ($y,x$) central plane is plotted in (\textit{c}), whereas $N^*(v)$ is plotted in (\textit{d}).
The sample extent is similar to Figure \ref{fig:Fig3a}.
}
\label{fig:Fig4a}
\end{figure}

\begin{figure}
\centering
\includegraphics[width=\textwidth]{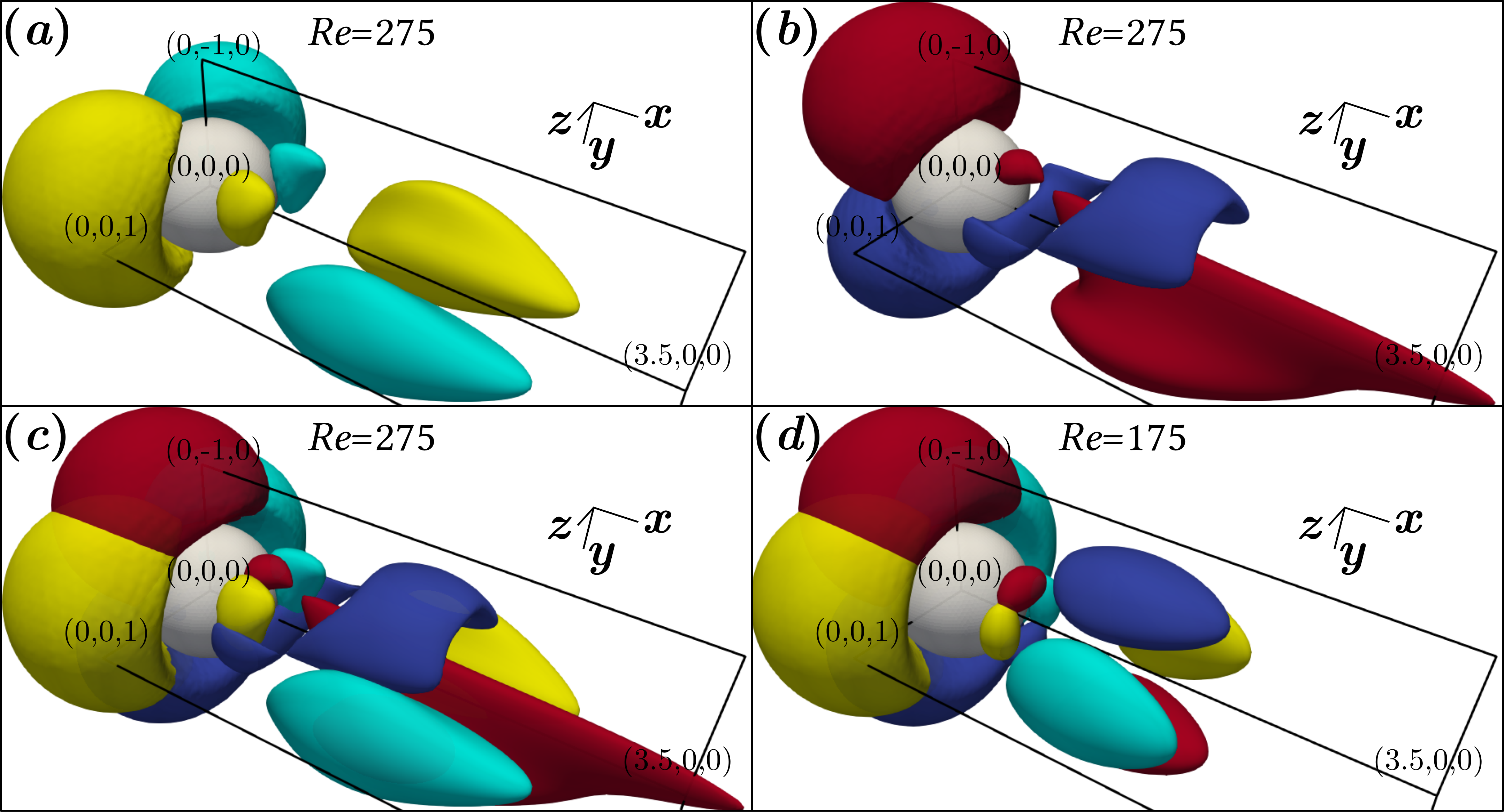}
\caption{Three dimensional spatial structure of velocity components,  $v$ and $w$. The surface contours of $w=-0.06$ and $0.06$ are plotted respectively in cyan and yellow in (\textit{a}), and $v=-0.06$ and $0.06$ contours are plotted respectively in blue and red in (\textit{b}).
(\textit{c}) and (\textit{d}) present both the $v$ and $w$ contours for the oblique $Re=275$ and axisymmetric $Re=175$ flow fields respectively.
}
\label{fig:Fig5a}
\end{figure}

Figure \ref{fig:Fig4a} presents the spatial distribution of pressure $p$ and transversal component of velocity $v$ along the orthogonal ($y,x$) plane for various $Re$.
In the axisymmetric regime, in Figure \ref{fig:Fig4a}(\textit{a}), the modulus of $v$ is symmetric across the $y=0$ plane but not the $v$.
Similarly the modulus of $w$ is also symmetric across the $z=0$ plane in the axisymmetric regime, but not $w$.
Complexity arises in the oblique regime, as neither $p$ nor the modulus of $v$ remains symmetric in the Figure \ref{fig:Fig4a}(\textit{b}).
This is also seen in the population density distribution of $v$ in Figure \ref{fig:Fig4a}(\textit{d}), where the positive magnitudes of $v$ show dominance similar to Figure \ref{fig:Fig4a}(\textit{b}).
In contrast to $v$, the positive and negative magnitudes of $p$ are rather concentrated near the sphere respectively in the upstream and the downstream zones as in Figure \ref{fig:Fig4a}(\textit{a,b}).
Similar to Figure \ref{fig:Fig3a}(\textit{c,d}), the $N(p)$ of the upstream zone ($p\ge0$ population) does not show significant variability with $Re$ and exhibits a Lorentzian distribution.
The $N(p)$ of the downstream zone however shows local peaks at around $p=-0.1$, which marks the discontinuity in the sample population in Figure \ref{fig:Fig4a}(\textit{a,b}).
A three dimensional spatial structure of the velocity components $v$ and $w$ for the oblique $Re=275$ and the axisymmetric $Re=175$ cases are shown in Figure \ref{fig:Fig5a}, where the complexity in the oblique wake flow structure can be appreciated.

\begin{figure}
\centering
\includegraphics[width=\textwidth]{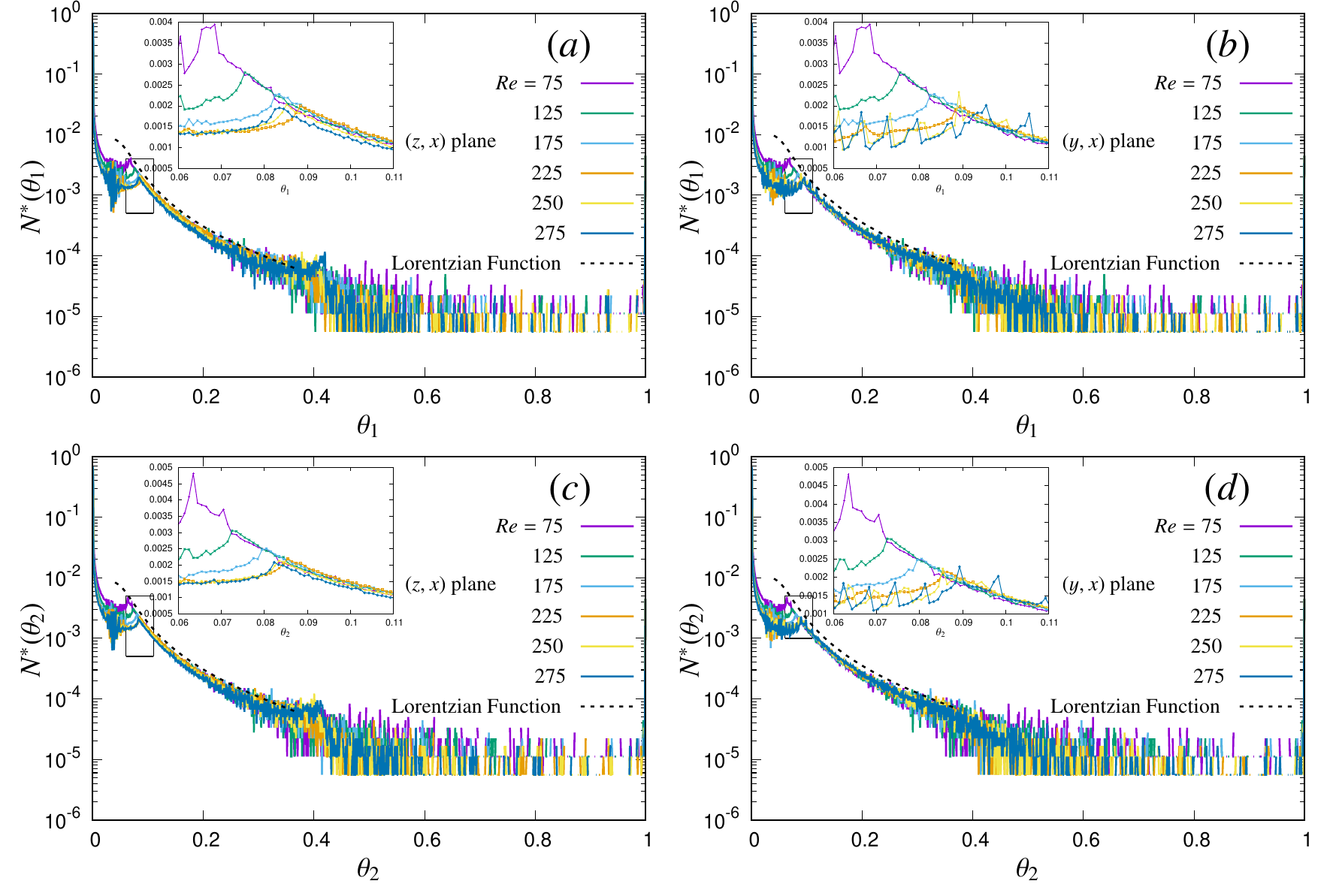}
\caption{Spatial evolution of the normalized population density of scalar $N^*(\theta_1)$ along the ($z,x$) plane is presented in (\textit{a}) and along the ($y,x$) plane in (\textit{b}).
Evolution $N^*(\theta_2)$ along the ($z,x$) plane is plotted in (\textit{c}) and along the ($y,x$) plane in (\textit{d}).
These orthogonal planes pass through the center of the sphere and extends to the entire simulated domain of [-3.5:3.5] in the horizontal $y,z$, and [-5:20] in the streamwise $x$ directions.
}
\label{fig:Fig6a}
\end{figure}

Figure \ref{fig:Fig6a} presents the population density distribution of the scalar fields $N(\theta_1)$ and $N(\theta_2)$ across two central orthogonal planes ($z,x$) and ($y,x$) (similar to the previous Figures).
Since the boundary conditions for the dimensionless scalars have a zero value in the ambient and a unit value on the sphere surface, their population density distribution shows the highest population around zero in Figure \ref{fig:Fig6a}, followed by a domain induced crescent zone, and then a Lorentzian distribution in the intermediate values gradually approaching the surface unit value.
The Lorentzian trend is again visible in the scalar population density, due to the similitude of the advection-diffusion equation of the scalars to the dynamics of momentum in regions with small pressure gradients. 
In the upstream region, the behaviour of velocity and scalars is very different due to the strong pressure gradient, while in the downstream region the difference is much milder.
A closer look to the density distribution in the insets shows that the steady axisymmetric cases do not show a well distinguishable difference in the number density at different scalar magnitudes with the increase in $Re$, but only the threshold magnitude for the start of the Lorentzian trend increases.
The shift in the threshold of Lorentzian trend is attributed due to the finite and a similar size of the simulation domain for all the $Re$ cases and due to the shrink in the lateral extent of the wake but a stretch in the streamwise direction with increasing $Re$.
The decrease in the sample population for the oblique cases in the left panel of Figure \ref{fig:Fig6a} for the orthogonal ($z,x$) plane is however due to the out of plane tilt of the wake which reduces the sample population.
Whereas in the right panel for the orthogonal ($y,x$) plane, we see a step-wise perturbation on top of an overall Lorentzian trend in the oblique wake regime as a result of its tilt in this plane.

\begin{figure}
\centering
\includegraphics[width=\textwidth]{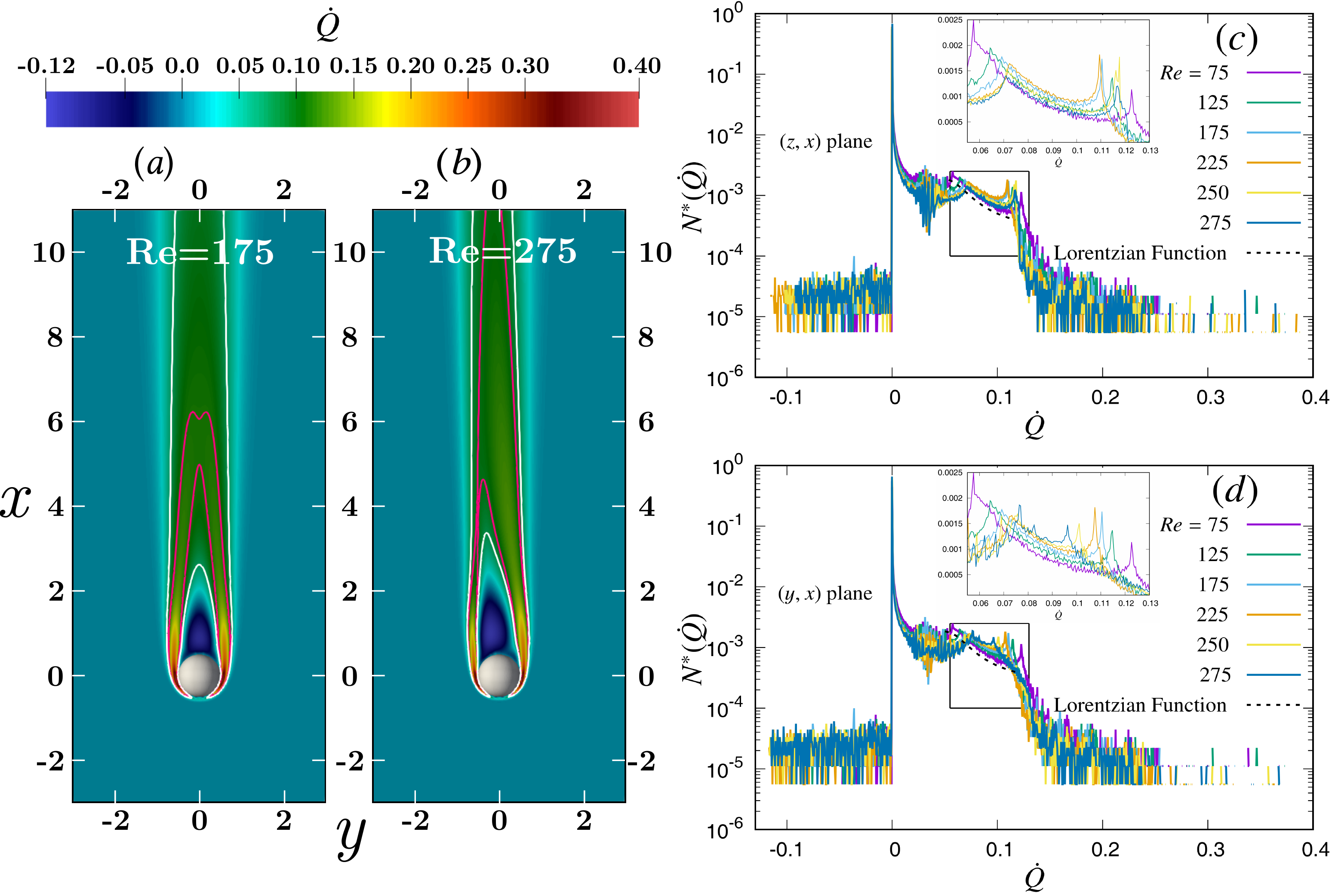}
\caption{Spatial distribution of convective scalar flux $\dot{Q}=u\cdot \theta_1$ for various $Re$. Spatial distribution of $\dot{Q}$ in color along the orthogonal ($y,x$) plane for the axisymmetric $Re=175$ in (\textit{a}) and for the oblique $Re=275$ in (\textit{b}).
The white contour lines represent $\dot{Q}=0.069$ in (\textit{a}) and $\dot{Q}=0.077$ in (\textit{b}), while the pink contour lines are at $\dot{Q}=0.11$ in (\textit{a}) and $\dot{Q}=0.096$ in (\textit{b}) respectively.
Normalized population density of pressure as $N^*(\dot{Q})$ across the entire orthogonal ($z,x$) and ($y,x$) central planes are plotted respectively in (\textit{c}) and (\textit{d}). The sample extent is similar to Figure \ref{fig:Fig3a}.
}
\label{fig:Fig7a}
\end{figure}

Figure \ref{fig:Fig7a} presents the spatial distribution of the convective scalar flux $\dot{Q}$ in the streamwise direction $x$, which is a product between $\theta$ and $u$. 
Spatial distribution of $\dot{Q}$ along the orthogonal ($y,x$) plane in Figure \ref{fig:Fig7a}(\textit{a,b}) is someway different from the other flow quantities, since it shows highest positive $\dot{Q}$ in the boundary layers and a negative $\dot{Q}$ in the recirculating zone due to negative $u$.
The non-symmetric spatial structure of the oblique ($Re=275$) scalar flux is visible in Figure \ref{fig:Fig7a}(\textit{b}).
The population density distribution $N(\dot{Q})$ along the orthogonal ($z,x$) and ($y,x$) planes shows a different structure as expected.
A Lorentzian trend is observed for a few limited sample populations, for example, for the samples between the white and pink contour lines in Figure \ref{fig:Fig7a}(\textit{a}) and (\textit{b}) respectively for $Re=175$ and 275.
These two contour lines correspond to the $\dot{Q}$ magnitudes from Figure \ref{fig:Fig7a}(\textit{d}) marking the beginning and the end of the Lorentzian trend for each individual $Re$.
Overall an increase in the sample population of $\dot{Q}$ is observed with increasing $Re$ within the zone with Lorentzian distribution.

\section{Discussions and Concluding Remarks} \label{Conclusion}

We present a detailed numerical analysis on the spatial structure of the wake flow using population density distribution for various Reynolds number in the steady wake regime. Incompressible Navier-Stokes equation is solved for the flow velocity and the one-way coupled advection-diffusion equations are solved for the scalars using the lattice Boltzmann method.
The spatial evolution of various flow quantities, such as, longitudinal velocity component $u$, pressure $p$, passive scalar $\theta$, convective scalar flux $\dot{Q}$ in the wake of steady axisymmetric regime ($Re\le220$) and oblique regime ($225\le Re\le 285$) using a population distribution function $N$, shows a Lorentzian distribution which is proportional to the inverse of the square of the flow quantity (for example, $N(p)\propto p^{-2}$).
This Lorentzian trend exhibits an algebraic decay in the number density of populations with different magnitudes of fluid quantities from the external ambient to the boundary layer in the wake and dominates the spatial distribution of the flow quantities outside the recirculating region.
The horizontal components of fluid velocity, $v$ and $w$, whereas show  different spatial distributions not attributable to a Lorentzian one.
Transition to the oblique wake regime at $Re\ge225$ in our simulations shows a complex three dimensional spatial evolution of the flow quantities, which also shows a Lorentzian trend.
The population density distribution for the longitudinal velocity component $u$, shows an increase in its number density with increasing $Re$.
Whereas the number density of the scalar populations remains the same for various axisymmetric $Re$.
This feature however changes in case of the convective scalar flux, where an increase in its number density is observed again with an increase in $Re$.

Quantification of scalar transport in the wake of spherical objects is important for understanding various physical phenomena. For example, Bhowmick et al. (2020) \cite{Bhowmick2020}, and also in Chouippe et al. (2019) \cite{Chouippe2019} and Krayer et al. (2020) \cite{Chouippe2020}, scalar transport in the wake is used to understand the spatial distribution of supersaturation in the wake of precipitating cloud hydrometeors.
By scaling the passive scalars as the temperature and the water vapor density fields around the droplets, we used three dimensional population distribution of supersaturation also in Bhowmick et al. (2020) \cite{Bhowmick2020} to quantify the supersaturated volume produced in the wake of precipitating cloud hydrometeors in presence of a sufficient temperature gradient in a slightly subsaturated cloudy ambient, which can activate cloud aerosols and thus contribute to the cloud life cycle.

\section*{Author Contributions}
Conceptualization, T.B., Y.W., M.I, G.B and E.B.; methodology, T.B., Y.W., M.I, G.B and E.B.; software, T.B. and Y.W.; simulation, T.B.; investigation, T.B., Y.W., M.I, G.B and E.B.; visualization, T.B., Y.W., M.I and G.B; writing-original draft, T.B., Y.W., M.I, G.B and E.B.; writing-review and editing, T.B., Y.W., M.I, G.B and E.B.; supervision, Y.W., M.I, G.B and E.B.; project administration, Y.W., G.B and E.B.; funding acquisition, E.B. All authors have read and agreed to the published version of the manuscript.

\section*{Funding}
This research was funded by the Marie - Sk\l odowska Curie Actions (MSCA) under the European Union's Horizon 2020 research and innovation programme (grant agreement no. 675675), and an extension to programme COMPLETE by Department of Applied Science and Technology, Politecnico di Torino.

\section*{Acknowledgments}
Scientific activities are carried out in Max Planck Institute for Dynamics and Self-Organization (MPIDS) and computational resources from HPC@MPIDS are gratefully acknowledged.
First author wishes to acknowledge Giuliana Donini, Guido Saracco, Mario Trigiante and Paolo Fino for support.

\section*{Conflicts of Interest}
The authors declare no conflict of interest. The funding sponsors had no role in the design of the study; in the collection, analyses, or interpretation of data; in the writing of the manuscript, or in the decision to publish the results.
 
\section*{References}
\bibliography{Symmetry2020.bib}

\end{document}